%% file: main.tex
\documentclass[sigconf]{acmart}

\usepackage{blindtext, graphicx}
\graphicspath{{images/}}
\usepackage[justification=centering]{caption}

\usepackage{amsmath}

\usepackage{booktabs} 
\usepackage[utf8]{inputenc}
\usepackage[english]{babel}
\usepackage{tabularx} 
\usepackage{multirow}
\usepackage{xcolor}
 
\usepackage{url}

\renewcommand{\arraystretch}{1.1}
\newcommand{\nlcell}[3][1.0]{\def\arraystretch{#1}\begin{tabular}[c]{@{}#2@{}}#3\end{tabular}}
\acmBooktitle{Preceedings of the ACM Mining Software Repositories conference}

\begin{document}
\title{Predicting Developers' IDE Commands with Machine Learning}

\copyrightyear{2018}
\acmYear{2018}
\setcopyright{acmcopyright}
\acmConference[MSR '18]{MSR '18: 15th International Conference on Mining Software Repositories }{May 28--29, 2018}{Gothenburg, Sweden}
\acmBooktitle{MSR '18: MSR '18: 15th International Conference on Mining Software Repositories , May 28--29, 2018, Gothenburg, Sweden}
\acmPrice{15.00}
\acmDOI{10.1145/3196398.3196459}
\acmISBN{978-1-4503-5716-6/18/05}


\author{Tyson Bulmer, Lloyd Montgomery, Daniela Damian}
\affiliation{
  \institution{University of Victoria}
  \city{Victoria}
  \country{Canada}
}
\email{{tysonbul, lloydrm, danielad}@uvic.ca}





\begin{abstract}
When a developer is writing code they are usually focused and in a state-of-mind which some refer to as flow. Breaking out of this flow can cause the developer to lose their train of thought and have to start their thought process from the beginning. This loss of thought can be caused by interruptions and sometimes slow IDE interactions. Predictive functionality has been harnessed in user applications to speed up load times, such as in Google Chrome's browser which has a feature called ``Predicting Network Actions". This will pre-load web-pages that the user is most likely to click through. This mitigates the interruption that load times can introduce. In this paper we seek to make the first step towards predicting user commands in the IDE. Using the MSR 2018 Challenge Data of over 3000 developer session and over 10 million recorded events, we analyze and cleanse the data to be parsed into event series, which can then be used to train a variety of machine learning models, including a neural network, to predict user induced commands. Our highest performing model is able to obtain a 5 cross-fold validation prediction accuracy of 64\%.
\end{abstract}

%
%



\maketitle

\input{sections/abstract}

\input{sections/introduction}

\input{sections/methodology}
\input{sections/results}

\input{sections/threats}
\input{sections/conclusion}

\bibliographystyle{ACM-Reference-Format}
\bibliography{main}

\end{document}

%% file: sections/abstract.tex
\begin{abstract}
When a developer is writing code they are usually focused and in a state-of-mind which some refer to as flow. Breaking out of this flow can cause the developer to lose their train of thought and have to start their thought process from the beginning. This loss of thought can occur for many reasons, one of which is slow IDE interactions. A potential technique for mitigating these slow IDE interactions is pre-loading interactions the user is likely to induce. Predictive functionality has been previously harnessed in user applications to speed up load times, such as in Google Chrome which has a feature called ``Predicting Network Actions''. This will pre-load web-pages that the user is most likely to click through. This helps mitigates the interruption that load times can introduce. In this paper we seek to make the first step towards predicting user commands in the IDE as a first step towards pre-loading IDE interactions. Using the MSR 2018 Challenge Data of over 3000 developer session and over 10 million recorded events, we train a variety of machine learning models---including a neural network---to predict user induced commands. Our highest performing model is able to obtain a 5-fold cross-validation prediction accuracy of 70\%.

\end{abstract}

%% file: sections/introduction.tex
\section{Introduction and Background}

The task of developing software is a thought-intensive process and interruptions can derail a train of thought easily. These interruptions can come in a variety of ways, one of which is slow loading time within an integrated development environment (IDE). Past research in bug prediction has touched on the topic of developer ``focus'' as a contributor to bug proneness. Di Nucci et al. propose that the higher a developer's focus on their activities, the less likely they are to introduce bugs to the code~\cite{DiNucci2015}. Zhang et al. proposed that certain file editing patterns by developers writing code could contribute to bug proneness as well. Two of their patterns which highlight the ``focus'' of a developer are the ``Interrupted'' and ``Extended'' editing patterns~\cite{Zhang2014}, which they found introduced 2.1 and 2.28  (respectively) times more future bugs than without.

The bigger picture of \textit{developer activity within an IDE}, as characterized by our own investigation of this research area, can be grouped into four categories: tracking, characterizing, enhancing, and predicting. Tracking developer activity is the most published area, due in large part to the necessity of tracking information before it can be characterized, enhanced, or predicted. One of the first published works in this area is by Teitelman and Masinter~\cite{Teitelman1981} who sought to expand the functionality of IDEs with automatic analyzing and cross-referencing of user programs. Takada et al. built a tool that automatically tracks and categorizes developer activity through the recording of keystrokes~\cite{Takada1994}. Their work introduced the concept of live-tracking developers, making way for the refined IDE activity tracking tools to follow.

Mylyn is a task and application lifecycle management framework for Eclipse \footnote{\url{https://www.eclipse.org/mylyn/}}. It began as Mylar, a tool for tracking developer activities in IDEs, focusing on building degree of interest (DOI) models to offer suggestions to developers on which files they should be viewing~\cite{Kersten2005}. Mylyn has always had a focus on tracking, characterizing, and enhancing developer activity, but not predicting~\cite{Murphy2006, Kersten2006, Fritz2007}.

DFlow is a tool developed by Minelli and Lanza to track and characterize developer activities within IDEs~\cite{Minelli2013, Minelli2015}. Their research focused on the tracking and characterization of developer activities, and not enhancing or predicting their activities.

Other notable IDE integrations includes Blaze---a developer activity enhancer by Singh et al.~\cite{Singh2014}, WatchDog---an IDE plugin that listens to UI events related to developer behavior designed to track and characterize developer activity by Beller et al.~\cite{Beller2015}, and FeedBaG---the developer activity tracker for Visual Studio \footnote{\url{https://www.visualstudio.com/}} coupled with research designed to track and characterize developer activity.

In this paper we use a corpus of over 10 million developer activity events generated in Visual Studio and recorded by FeedBaG~\cite{msr18challenge}, to train a model to predict developer commands. We formally address our intent with the following research question:

\begin{itemize}
    \item \textbf{RQ}: Can we use machine learning models to predict developer induced IDE commands?
\end{itemize}

Similar work has been done on trying to predict developer behaviour in the IDE by Damevski et al., where they abstracted recorded IDE event sequences using topic models to predict developer future tasks, such as debugging or testing~\cite{Damevski2017}. Our research differs from Damevski's work by harnessing a neural network and the granularity in which our model predicts command-level actions.

We believe that if we can predict user commands, then we can reduce slow command process times of actions by preprocessing the predicted user commands, similar to applications such as Google Chrome which has predictive page loading to speed up browsing \footnote{\url{https://support.google.com/chrome/answer/1385029}}.

This paper is structured as follows: Section \ref{sec:methodology} describes the methodology, Section \ref{sec:results} covers the results, Section \ref{sec:threats} discusses the threats to validity, and Section \ref{sec:conclusion} concludes the paper.

%% file: sections/methodology.tex
\section{Methodology}
\label{sec:methodology}

In this section we describe our process to analyze, clean, and format the data into a form which can represent features and labels for our machine learning models, as well as our selection of models, and their performance evaluation.

\subsection{Data Description}
The dataset we use is supplied by Proksch et al.~\cite{msr18challenge} for the 2018 MSR Challenge\footnote{https://2018.msrconf.org/track/msr-2018-Mining-Challenge}. The dataset was collected by FeedBaG++\footnote{http://www.kave.cc/feedbag}, an IDE integration that logs events occurring within the IDE including developer activities, IDE environment changes, navigation, and more. Released in March 2017, this dataset contains over 10 million events, originating from 81 developers and covering a total of 1,527 aggregated days of developer work.

\begin{figure}
    \centering
    \includegraphics[width=\columnwidth]{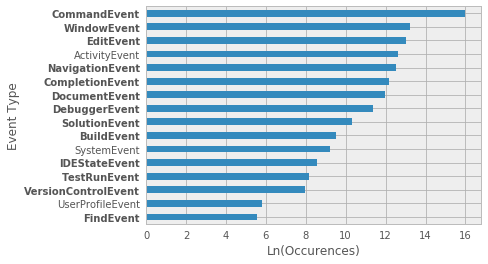}
    \caption{Frequency (ln()) of event-type occurrences in sessions, with utilized event-types in bold}
    \label{fig:event-type-frequency}
\end{figure}

The dataset contains 18 unique IDE event types, but we only utilized 13 event types initiated by the developer. The most common developer-induced events are ``Command'' events which are an ``invocation of an action'' by the developer, e.g. ``clicking a menu button or invoking a command via shortcut''~\cite{msr18challenge}.
A complete listing and descriptions of these events can be found on the data supplier's website\footnote{http://www.kave.cc/feedbag/event-generation}. Figure \ref{fig:event-type-frequency} shows the natural logarithm of the distribution of 16 of the event types across all sessions, with the 13 developer-induced events in bold (2 of the 18 types were lost in the original parsing of the data). From this we can see that the Command event type is the most commonly occurring event type.

\subsection{Data Processing}
\label{ssec:processing}
In this section we describe the steps taken to select, cleanse, and format the dataset. The dataset began with \textasciitilde10.7 million events.

\subsubsection{Event-Type Selection}
As previously mentioned, we only utilize 13 of the original 18 event types available in the dataset, which are developer-induced. These 13 event types are desired because they are invoked by the developer, and therefore part of the developer's working pattern. Although other dynamic system responses are occurring, we wanted to focus specifically on the workflow of developers. The non-developer-induced event types, ``SystemEvent'', ``InfoEvent'', and ``ErrorEvent'', as well as the ``ActivityEvent'' (recorded when the mouse moves) and ``UserProfileEvent'' (recorded when user profile for the FeedBaG++ tool is modified), were removed from the dataset. Roughly 300,000 events were removed due to this selection.

\subsubsection{Data Cleansing}
Just over 400,000 events had been recorded twice with the same triggered-at date, event type and session ID. These events are considered duplicate records of the same triggered event and therefore were removed from the dataset.

From our manual investigation of the data, we found that many events, including Command events, were repeated hundreds of times in quick succession, milliseconds apart. This behaviour is not easily explained, nor is it insignificant, so we did not want to remove them completely. However, we believed that this repetition of events would negatively influence our models as they would learn to predict these heavily repeated events and still achieve high performance in the test environment---while failing to perform well in the real environment. To avoid this situation, while preserving the existence of the repetitions, we applied the regular expression notation of ``+'' (meaning one or more) to events which repeated one or more times. For example, the sequence ``Command, Command, Command'' would be replaced with ``Command+.'' This conversion process removed just over 6 million events.

\subsubsection{Event Representation}
Most of the event types have additional descriptors about the recorded event which distinguish them from other similar event types. For example, the Window Event also records an ``action'' value which represents whether the action of the Window event was to move, open, or close the window. This is important data for helping understand the more precise actions of the developer, and this should be captured by our models. For each event type, we look through the supplied descriptors and map the associated values which are generated by the event.

We append onto the event type the respective descriptor. For example, a Command event instance would be represented as follows: \textit{``CommandEvent-Refresh''}, where \textit{``CommandEvent''} is the event type and \textit{``Refresh''} is the additional descriptor.

\subsection{Feature and Target Class Extraction}
Due to the \textit{event stream} structure of the data, it is important to frame how we are extracting the features and target classes from the data. 

Although our data contains 13 developer-induced event types, we are only trying to predict Command events with their additional descriptors. There are a total of 651 unique Command descriptors, but since the 61 most frequently used Command descriptors account for 90\% of the total Commands invoked, we narrowed down the target classes to just those 61. In trying to predict developer-induced events (specifically Commands) for the purpose of preprocessing them, focusing on the most frequent Commands descriptors allows us to target the most likely actions that will occur.

Secondly, the data is recorded as multiple sessions of user event streams, and we need to frame what the features of the target classes are. For illustrative purposes, let's say one session's event stream is \textit{G1, G2, C1, G3, C2}, where \textit{C} events are one of the 61 Command descriptor types described above, and \textit{G} events are any other generic event, including any of the other command descriptor types, or any of the other event types along with their various additional descriptors. As described above, we are only interested in predicting the top 61 Command descriptors, so from the example we would extract two rows in our X and y machine learning matrices (where X contains the features and y contains the target classes). The first row of features would be \textit{\{G1, G2\}} with the target class of \textit{C1}, and the second row of features would be \textit{\{G1, G2, C1, G3}\} with the target class of \textit{C2}.

\subsection{Models}
We used four functionally different models to explore which would perform the best for the given task of predicting the next event. We trained and tested on a randomly sampled 100,000 event series from our 3.9 million events (remaining after processing described in Section \ref{ssec:processing}). This reduced sample is due to the Logistic Regression model's computationally expensive algorithm performed with over 12 thousand features which are a result of the largest N-Gram range. For all the models, we conduct 5-fold cross-validation, this was chosen in favour of 10-fold cross-validation, as it required training only half as many models which reduces the computational load. The models we tested are as follows:

\subsubsection{Naive Bayes}

We harnessed Sci-Kit-learn's implementations of the classification models Bernoulli and Multinomial Naive Bayes which are based on conditional probability and used commonly for text classification tasks~\cite{murphy2006naive}. To represent these series of events we use a method called N-Grams~\cite{lodhi2002text} which takes series of tokens, in our case \textit{events}, and breaks them up into tuples of N length. Take the following event series for example:
\textit{E1, E2, E3}, A bigram breakdown (2-Gram) of it would look like (\textit{E1},\textit{E2},) (\textit{E2},\textit{E3}.) These N-Grams can be a variety of length which can affect the performance of our models so we experiment with runs on different N-Grams lengths ranging from 1 to 3. We limit the max N-Gram range to 3 due to the limited computational resources available. Once we have these N-Gram representations of our features, we put them through a count vectorizer, which produces a matrix that represents the counts of N-Grams. 

\subsubsection{Logistic Regression}
To widen our variety of models, we also tried Scikit-learn's Logistic Regression model~\cite{lee2006efficient}. Similarly to the Naive Bayes models, vectorized N-Grams are used to represent the data.

\subsubsection{Neural Network}
We also implemented a Neural Network (NN), using the high-level neural networks API Keras, which consisted of rectified linear unit hidden layers of 500, and 100 nodes, including a dropout layer with a dropout rate of 0.5 between the two hidden layers to help reduce overfitting~\cite{srivastava2014dropout}, and then a final layer with softmax activation for the step of classification. The way in which we represent the data for this model requires that we encode the events to categorical numbers, and that we one-hot encode the target classes, which is a method for representing multiple classes as binary features~\cite{862024}.

\subsection{Reporting Results}
To compare the performances of our models, which are performing multi-class classification, we use two measures: accuracy and micro-averaged receiver operator characteristic area under the curve (ROC AUC). The accuracy measure is the total accuracy across the 5 folds of the 5-fold cross validation. Meaning, the predictions from each fold across the entire dataset are aggregated and compared to the actual classes. Since our target classes are imbalanced, we use the micro-averaged ROC AUC, which is the weighted average ROC AUC for each class~\cite{Yang1999}. Meaning, the ROC AUC was calculated for each class as if it was a binary classification task, and then we take the weighted average of these ROC AUCs to account for the disproportion of class data points. The micro-average is an aggregate of the contributions of all classes to compute the average metric. For both the accuracy and micro-averaged ROC AUC, higher values are better.

%% file: sections/results.tex
\section{Results}
\label{sec:results}

In this section we discuss the results of each of our predictive models by reporting the accuracy and micro-averaged AUC ROC of the 5-fold cross-validation results.

\subsection{Naive Bayes}

The Bernoulli Naive Bayes models achieved its highest accuracy of 20.52\% with N-Grams of ranges 1, 2, and 3. Its micro-average ROC AUC was 0.60, as shown in Table \ref{table:all-results}. AUC's for each target class, or target command, ranged from 0.5 to 0.96, with a mean of 0.55 and standard deviation of 0.10.


The Multinomial Naive Bayes models achieved its highest accuracy of 20.14\% with N-Grams of ranges 1, 2, and 3. Its micro-average ROC AUC was 0.59, as shown in Table \ref{table:all-results}. AUC's for each target class, or target command, ranged from 0.5 to 0.97, with a mean of 0.61 and standard deviation of 0.14.


\subsection{Logistic Regression}

The Logistic Regression model achieved its highest accuracy of 35.87\% with N-Grams of ranges 1, 2. Its micro-average ROC AUC was 0.67, as shown in Table \ref{table:all-results}. AUC's for each target class, or target command, ranged from 0.50 to 0.95, with a mean of 0.63 and standard deviation +/- 0.12.


\subsection{Neural Network}

The NN performed achieved the highest accuracy among the classifiers with an accuracy of 64.39\%. Its micro-average ROC AUC was 0.82, as shown in Table \ref{table:all-results}. AUC's for each target class, or target command, ranged from 0.5 to 0.98, with a mean of 0.72 and standard deviation +/- 0.17.

\input{sections/includes/allmodelstable}

%% file: sections/includes/allmodelstable.tex
\begin{table}
\centering
\caption{All models results}
\label{table:all-results}
\resizebox{\columnwidth}{!}{
\begin{tabular}{|c|c|l|l|}
    \hline
    Model   & \nlcell{c}{N-Gram\\Range} & Accuracy (5-fold) & Micro AUC      \\ \hline
        \multirow{3}{*}{\nlcell{c}{Bernoulli\\Naive Bayes}}
                & {[}1,1{]}     & 15.74\%    & 0.57     \\ \cline{2-4}
                & {[}1,2{]}     & 18.77\%    & 0.57     \\ \cline{2-4}
                & {[}1,3{]}     & 20.53\%    & 0.60     \\ \hline
        \multirow{3}{*}{\nlcell{c}{Multinomial\\Naive Bayes}}
                & {[}1,1{]}     & 16.31\%    & 0.57     \\ \cline{2-4}
                & {[}1,2{]}     & 17.73\%    & 0.58     \\ \cline{2-4}
                & {[}1,3{]}     & 20.14\%    & 0.59     \\ \hline
        \multirow{3}{*}{\nlcell{c}{Logistic\\Regression}}
                & {[}1,1{]}     & 25.37\%    & 0.62     \\ \cline{2-4}
                & {[}1,2{]}     & 35.88\%    & 0.67     \\ \cline{2-4}
                & {[}1,3{]}     & 35.22\%    & 0.67     \\ \hline
        \multirow{1}{*}{Neural Network}
                & ---           & \textbf{64.39\%}  & \textbf{0.82}  \\ \hline
\end{tabular}}
\end{table}

%% file: sections/threats.tex
\section{Threats to Validity}
\label{sec:threats}

This section discusses the construct, internal, and external threats to the validity of this study and their respective mitigations.

\textbf{Construct Validity.} The main concern regarding construct validity in this research is the assumption placed on the data received. We only have a small description of what the developers experience and projects were, meaning they could have been working on anything. Some of their IDE activities might have not even been code related, or their IDE could have been open while using other applications and it still would have recorded various activities like those that pertain to ``Window'' events. However, we mitigated this by only predicting developer-induced events, which are more likely to be done during active development.

\textbf{Internal Validity.} A threat to internal validity in this study comes with the data source and its cleaning, as the data started in a JSON format and was converted to a SQL schema. The data has many other attributes which were available, but not able to be used for this study. However, since there is always the opportunity for more data to be used in a machine learning implementation to improve results, we see the impact of excluding that data as minimal to the study design.

\textbf{External Validity.} A threat to external validity in this study pertains to the events and commands recorded by the FeedBaG++ tool. However, the recorded event types we utilized are mostly generic, such as TestRunEvent and BuildEvent, and therefore this threat is mitigated by the general nature these attributes.

%% file: sections/conclusion.tex
\section{Conclusion and Future Work}
\label{sec:conclusion}

To help keep developers focused, and therefore less likely to introduce bugs, one could ensure that their IDEs perform with little to no delay time. A step towards ensuring this can be to harness predictive models combined with the recently released data of recorded developer activity events. Our findings suggest that a Neural Network can be used to accomplish this task with an accuracy of 64\%. 


\vspace{3mm}

\noindent\fbox{
    \parbox{.455\textwidth}{
        \textbf{Answer to RQ:} Based on our results, we see that it is possible to achieve an accuracy of 64\% across a target set of 61 possibilities while maintaining a micro-averaged ROC AUC of 0.82; therefore, we conclude that the answer to our research question is yes. 
    }
}

\vspace{3mm}


Future work which can harness these results would be to integrate command prediction into an IDE to test its feasibility and actual performance in a development environment. For improved prediction performance, given access to more computational resources, one could try using a Recurrent Neural Network (RNN) with the entire (unfiltered) event series. As RNN's are able to achieve high performance when used for time series prediction tasks~\cite{husken2003recurrent}.


%